\documentclass[a4paper,11pt,leqno]{article}
 \usepackage{epsfig, graphics, amssymb, amsfonts, euscript ,float,graphicx,color}
\def\today{\number\day\space\ifcase\month\or
January\or February\or March\or April\or May\or June\or July\or
August\or September\or October\or November\or December\fi
\space\number\year}

\makeatletter
\def\@begintheorem#1#2{\trivlist \item[\hskip \labelsep{\sc #1\ #2.}]\sl}
\def\@opargbegintheorem#1#2#3{\trivlist
\item[\hskip \labelsep{\sc #1\ #2\ (#3).}]\sl}
\def\@endtheorem{\endtrivlist}
\def\@sect#1#2#3#4#5#6[#7]#8{\ifnum #2>\c@secnumdepth
\let\@svsec\@empty\else
\refstepcounter{#1}\edef\@svsec{\csname the#1\endcsname.\hskip
0.5em}\fi \@tempskipa #5\relax \ifdim \@tempskipa>\z@
\begingroup #6\relax
\@hangfrom{\hskip #3\relax\@svsec}{\interlinepenalty \@M #8\par}
\endgroup \csname #1mark\endcsname{#7}\addcontentsline
{toc}{#1}{\ifnum #2>\c@secnumdepth \else
\protect\numberline{\csname the#1\endcsname}\fi #7}\else
\def\@svsechd{#6\hskip #3\relax\@svsec #8.\csname #1mark\endcsname
{#7}\addcontentsline {toc}{#1}{\ifnum #2>\c@secnumdepth \else
\protect\numberline{\csname the#1\endcsname}\fi #7}}\fi
\@xsect{#5}}
\makeatother 

\newcommand{\be}{\begin{equation}}
\newcommand{\bT}{\begin{array}}
\newcommand{\eT}{\end{array}}
\newcommand{\de}{\end{equation}}
\newcommand{\ee}{\end{equation}}

\def\t{\gamma}

\def\text#1{{\quad \hbox{#1} \quad}}

\def\ds{\displaystyle}
\def\sec{{\prime\prime}}

\begin{document}
\title{\bf On  similarity and pseudo-similarity solutions of  Falkner-Skan boundary layers}
\author{
 \bf M.~\textsc{\bf Guedda} and  \bf Z.~\textsc{\bf Hammouch}\footnote{Corresponding author. E-mail: zakia.hammouch@u-picardie.fr}  \\
{\footnotesize \it LAMFA, CNRS UMR 6140, Universit\'e de Picardie
Jules Verne,
}\\
{\footnotesize \it Facult\'e de Math\'ematiques et d'Informatique,
33, rue Saint-Leu 80039 Amiens, France} }
\date{}
\maketitle

\noindent \rule{17cm}{0.02cm} {\bf Abstract}\\

{\footnotesize The present work deals with the two-dimensional
incompressible, laminar, steady-state boundary layer equations.
First, we determine a family  of velocity distributions outside
the boundary layer such that these problems  may have  similarity
solutions. Then, we examen in detail new exact solutions, called
{\it Pseudo--similarity}, where the external velocity varies
inversely--linear with the distance along the surface $ (u_e(x) =
u_\infty x^{-1}). $  The analysis shows that solutions exist only
for a lateral suction. For specified conditions, we establish the
existence of  an infinite number of solutions, including monotonic
solutions and solutions which oscillate an infinite number of
times and tend to a certain limit. The properties of solutions
depend on
the suction parameter. Furthermore, making use of the fourth--order Runge--Kutta scheme together with the shooting method, numerical solutions are obtained.}\\
 \noindent{\scriptsize {\it \bf Keywords:}\quad
Similarity solutions, Pseudo-similarity solutions, Falkner-Skan
problem, Boundary Layer, Stretching surfaces.\newline
 }
\rule{17cm}{0.02cm}\\

\section{Introduction}\label{Sect.1}
\setcounter{equation}{0} \setcounter{theorem}{0}
\setcounter{lemma}{0} \setcounter{remark}{0}
\setcounter{corollary}{0}

\qquad In this paper we are concerned with the classical
two-dimensional laminar incompressible boundary layer flow past a
wedge or a flat plate \cite{Sch}. For the first approximation, the
model  is described by   the Prandlt equations or the boundary
layer equations
\begin{equation}\label{eq:Prandtl}
u\partial_x u + v\partial_y u= u_e\partial_x
u_e+\nu\partial^2_{yy} u, \quad \partial_x u + \partial_y v = 0,
\end{equation}
where $(x,y)$ denote the usual  orthogonal Cartesian coordinates
parallel and normal to the boundary $ y  = 0 $ (the wall),  $ u$
and $v $ are the corresponding
 velocity components,  and the constant $ \nu > 0 $ is the kinematic-viscosity.   The function $ u_e $  is a given exterior streaming velocity flow
 which is assumed throughout the paper to be nonnegative function of the single variable $x; u_e = u_e(x),$ and is such
that $ u(x,y) $ tends to $ u_e(x) $ as $ y\to \infty.$
 Equations (\ref{eq:Prandtl})  can be written in the form
\begin{equation}
\partial_y\psi\partial_{xy}^2\psi - \partial_{x}\psi\partial_{yy}^2\psi = u_e\partial_x u_e+\nu\partial^3_{yyy}\psi,
\end{equation}
where $ \psi $ is the well--known stream function defined by   $ u
= \ds\partial_{y}\psi, v = -\partial_{x}\psi.$

This problem with appropriate external velocity flow has been the
main focus of studies of particular exact solutions. Research on
this subject
 has a long history, which dates to the
pioneering works by  Blasius \cite{Blas} and Falkner and Skan
\cite{FS} in which the external velocity is given by
\begin{equation}
 u_e(x) = u_\infty x^m \quad (u_\infty > 0).\end{equation}
 Their investigations lead to solutions to (1.2) in the form
\begin{equation}\label{eq:similarity}
 \psi(x,y) = ax^{\alpha} f(byx^{-\beta}), \  a, b > 0,
\end{equation}
where
\begin{equation}
\alpha = \frac{m+1}{2}\ \mbox{and}\ \beta =
\frac{1-m}{2}.\end{equation} The equation for $ f $ is the
well-known equation obtained by Falkner-Skan \cite{FS}
\begin{equation}\label{eq:FS}
f^{\sec\prime}+ \frac{m+1}{2}ff^\sec + m({1-f^\prime}^2)=0\quad
\mbox{ on}\quad (0,\infty),
\end{equation}
or, if $ m > -1,$
\begin{equation}\label{eq:FSgamma}
f^{\sec\prime}+ ff^\sec + \sigma({1-f^\prime}^2)=0\quad \mbox{
on}\quad (0,\infty),
\end{equation}
where
\[ \sigma= \frac{2m}{m+1}.\]
Such equations occur at  wedge flows \cite[ p. 170]{Sch}. These
equations have received considerable attention in the literature.
We refer the reader to the works of Rosenhead \cite{Rose},
Schlichting and Gersten \cite{Sch}, Weyl \cite{Weyl} and Coppel
\cite{Cop11} and the references therein. Note that from (1.4) it
is easily verified that,
\[ \psi(x,y) = \left(\frac{x}{x_0}\right)^{\alpha}\psi\left(x_0,y\left(\frac{x}{x_0}\right)^{-\beta}\right).\]
This means that a solution $ \psi(x,y) $ for $ y $ fixed is
similar to the solution $ \psi(x_0,y) $ at a certain $ x_0.$
 This
solution is called invariant or similarity solution and the
function $ f
$ is called the shape function or the dimensionless stream function.\\
The broad goals of this paper is  to study equation (1.7) when
taking the limit $\sigma \to -\infty.$ This limit case,
corresponding
 to $ m = -1,$ is that of flow in a two-dimensional divergent channel (or sink flow) \cite[p. 170]{Sch}. We prove    that transformation (1.4) is much too restrictive; that
is problem (1.1) has no  similarity solution and we shall see, by
rigouros arguments, that   the term $ \gamma log(x) $ must be
added to the expression (1.4) and the surface must be permeable
with suction to obtain new exact solutions which are not
similarity. In addition, we shall prove    that there is an
infinite number of solutions. These results are given in Section
3. Before this analysis, we shall identify, in Section 2, external
flows, such that problem (1.2) may have similarity solutions.  The
main result of this section  indicates that problem (1.2) has
solutions under the form (1.4) if the external flow is of the
power--law  type (1.3).

\section{Similarity solutions}
\setcounter{equation}{0} \setcounter{theorem}{0}
\setcounter{lemma}{0} \setcounter{remark}{0}
\setcounter{corollary}{0} \qquad In this section we shall obtain
external flows such that the partial differential equation (1.2)
accompanied by the boundary condition
\begin{equation}
 \lim_{y\to \infty}\partial_y\psi(x,y) = \partial_y\psi(x,\infty) = u_e(x),
\end{equation}
has a solution under the form (1.4), where $ \alpha+\beta = 1.$

The main  problems, arising in the study of similarity solutions,
are related to the existence of the exponents $ \alpha $ and $
\beta $  and to the rigorous study of the ordinary differential
equation satisfied by the profile $ f$ which is, in general,
nonlinear. For the layer equation (1.2), the classical approach
for identifying $ \alpha $ and $ \beta $ is the  scaling and
transformation group \cite{Bar}. The essential idea  is to seek $
a $ and $ b $ such that if $ \psi $ satisfies (1.2)   the new
function $ \psi_\kappa(x,y) = \kappa^a\psi(\kappa^b x,
\kappa y) $ is also a solution. \\
Let  $ \psi $ be a stream-function to (1.2) defined by (1.4) where
$ \alpha + \beta = 1.$ Assume first that $ \beta\not=0.$  We
choose $ a = -\frac{\alpha}{\beta},\ b = \frac{1}{\beta},$ and
define $\psi_\kappa(x,y) = \kappa^a\psi\left(\kappa^bx,\kappa
y\right).$ Hence $ a + b = 1, \psi\equiv \psi_\kappa $ and
\[ L(\psi_\kappa)(x,y) = \kappa^{a+3}L(\psi)(\kappa^bx,\kappa y )\] for any $ \kappa > 0,$
where $ L $ is the operator defined by
\[ L(\psi) =\partial_y\psi\partial_{xy}^2\psi - \partial_{x}\psi\partial_{yy}^2\psi-\nu\partial^3_{yyy}\psi.\]
According to equation (1.2) we deduce
\[ h(x)=\kappa^{a+3}h(\kappa^bx),\]
where $ h(x) =  u_e(x)\partial_x u_e(x).$ In particular, for fixed $
x_0 > 0 $
\[ h(\kappa^bx_0)=\kappa^{-(a+3)}h(x_0).\]
Setting  $ x = \kappa^bx_0 $ we infer
  \[h(x)=x^{-\frac{a+3}{b}}x_0^{\frac{a+3}{b}}h(x_0).\]
Solving the equation
\begin{equation} u_e\frac{d u_e}{dx} =  x^{-\frac{a+3}{b}}x_0^{\frac{a+3}{b}}h(x_0)\end{equation} yields us
\begin{equation}\label{eq:UE}
 u_e^2(x) = {c_1x^{2m}+c_2},\end{equation}
for all $ x > 0, $ where $  m = \alpha-\beta$ and $ c_1 $ and $
c_2 $ are constants, for $\beta\not=0$, since $ -\frac{a+3}{b}
+1=2(\alpha-\beta). $\\ For  $ \beta = 0, $ hence $ \alpha =m =
1,$  the new function
\[ \psi_\kappa(x,y)= \kappa^a\psi(\kappa^{-a}x,y),\]
for any fixed $ a \not= 0, $ is equivalent to $ \psi $ and
satisfies
\[ L(\psi_\kappa)(x,y) = \kappa^{a}L(\psi)(\kappa^{-a}x,y )\] for any $ \kappa > 0.$ Arguing as in  the case $ \beta \not=0 $ one arrives at (2.3) with $ m
= 1.$ Next, because $
\lim_{y\to\infty}x^{2m}f^\prime(yx^{-\beta})^2 = c_1x^{2m}+ c_2,$
the function $ {f^\prime}^2 $ has a finite limit at infinity,
which is unique and is given by $ c_1 + c_2x^{-2m}.$ This is
acceptable only for $ c_2 = 0.$

 The above result indicates, in
particular, that for a prescribed external velocity satisfying
(2.3)
 the real numbers
$ \alpha $ and $ \beta $  are given by (1.5) and  condition (2.3)
is necessary and sufficient to obtain similarity solution under
the form (1.4) where $ \alpha+\beta = 1.$  However, for a
 general external velocity, it is possible to obtain an exact solution which is not similarity solution as it is seen in
\cite{LCB}. In this paper the authors considered
 \[ u_e(x) = c_1x^{1/3} + c_2x^{-1/3}. \]
A stream--function $ \psi $ is given by
\begin{equation}
\psi(x,y)=x^{2/3} f(yx^{-1/3}) + cx^{-\frac{1}{3}}y,\quad c =
const,\end{equation} where $ f $ is a solution of
\begin{equation}
\nu f^{\sec\prime}\ +\ \alpha ff^\sec
-(\alpha-\beta){f^\prime}^2=\xi,\quad \xi = const.
\end{equation}

\section{ The pseudo-similarity  solutions}
\setcounter{equation}{0} \setcounter{theorem}{0}
\setcounter{lemma}{0} \setcounter{remark}{0}
\setcounter{corollary}{0} \qquad In the present section we
restrict the attention to the case $ m = -1 $  and get new
solutions to the problem
 \begin{equation}
\partial_y\psi\partial_{xy}^2\psi - \partial_{x}\psi\partial_{yy}^2\psi = \nu\partial_{yyy}^3\psi-u_\infty^2 x^{-3},
\end{equation}
subject to  the  boundary conditions
\begin{equation}
\partial_y\psi(x,0) = u_w x^{-1},\quad \partial_x\psi(x,0) = -v_wx^{-1},\quad \partial_y\psi(x,\infty)=u_\infty x^{-1},
\end{equation}
where  $ v_w $ is a real number ( $v_w>0$ for suction and $v_w<0$
for injection),    $ u_w $ and $ u_\infty $ are nonnegative and
satisfy $ u_w <  u_\infty.$  The velocity distribution $ u_e
=\frac{u_\infty}{x }$ is found in the case of divergent channel
(or sink flow) \cite[p. 170]{Sch}.  The analysis of this section
is motivated by the work by  Magyari, Pop and Keller \cite{MPK}
concerning a boundary--layer flow, over a permeable continuous
plane surface, where the $x$-component velocity tends to zero for
$ y $ large ($u_\infty = 0$). In \cite{MPK} the authors  showed
that if $  m = -1 $ problem (1.2),(2.1) has no solution in the
usual form (1.4). For $ u_\infty\not=0$ and according to Section
2, the function $ \psi $ can be written as
\[ \psi(x,y) = \sqrt{\nu u_\infty}f\left(\sqrt{\frac{u_\infty}{\nu}}yx^{-1}\right).\]
Since $ v_w = \frac{m+1}{2}\sqrt{\nu u_\infty}f(0) $ (see the
Appendix) we deduce  $ v_w = 0 $ for $ m=-1 $  and the following
oridinary differential equation for $ f $
\begin{equation}
\left\{\begin{array}{lll}
 f^{\sec\prime}  +{f^\prime}^2-1 = 0, \\
\\
f^\prime(0) = \zeta, f^\prime(\infty) = 1,
\end{array}\right.\end{equation}
where $ \zeta = \frac{u_w}{u_\infty} $ is in the interval $ [0,1)
$ and $ f(0) $ can be any real  number.  Since this problem does not
contain $ f $ it is convenient to study the equation satisfied by
$ \theta = f^\prime; $ that is
\begin{equation}
\left\{\begin{array}{lll}
 \theta^{\sec}  +{\theta}^2-1 = 0, \\
\\
\theta(0) = \zeta, \theta(\infty) = 1,
\end{array}\right.\end{equation}
 \\ The stability of equilibrium point  $ (1,0) $ of (3.4) cannot
be determined from the linearization. To analyze the behavior of
the nonlinear equation (3.4)$_1$, we observe that
\[ E^\prime(t)=0,\]
where $E$ is the Liapunov function given by\\
\[ E(t) = \frac{1}{2}\theta^\prime(t)^2 + \frac{1}{3}\theta(t)^3 - \theta(t).\]
Then, for  some  constant $ c,$ the following
\[ \theta^{\prime}=\pm\sqrt{2}\left(c+\theta-\frac{1}{3}\theta^3\right)^{1/2},\]
holds. The analysis of the algebraic equation of the phase path in
the phase plane reveals that the equilibrium point $(1,0)$ is a
center.
Hence, Problem (3.4) has no solution for any $ \zeta >-1 $ except the trivial one $ \theta = 1$ (see Fig. 3.1). \\
Note that if we impose the condition $ \theta(\infty) = -1 $
instead of $ \theta(\infty) = 1,$ which is also of physical
interest, it is easy to see that, for any $ \zeta \leq 2, $ there
exists a unique solution up to translation. This solution
satisfies
\[ \frac{1}{2}\theta^\prime(t)^2 + \frac{1}{3}\theta(t)^3 - \theta(t)= \frac{2}{3},\]
and we find that
\[ \theta(t) = 2-3\tanh^2\left[\pm t/\sqrt{2}+{\rm arctanh}\left\{(2-\zeta)/3)^{1/2}\right\} \right].\]

\begin{center}
\includegraphics[height=2.8in]{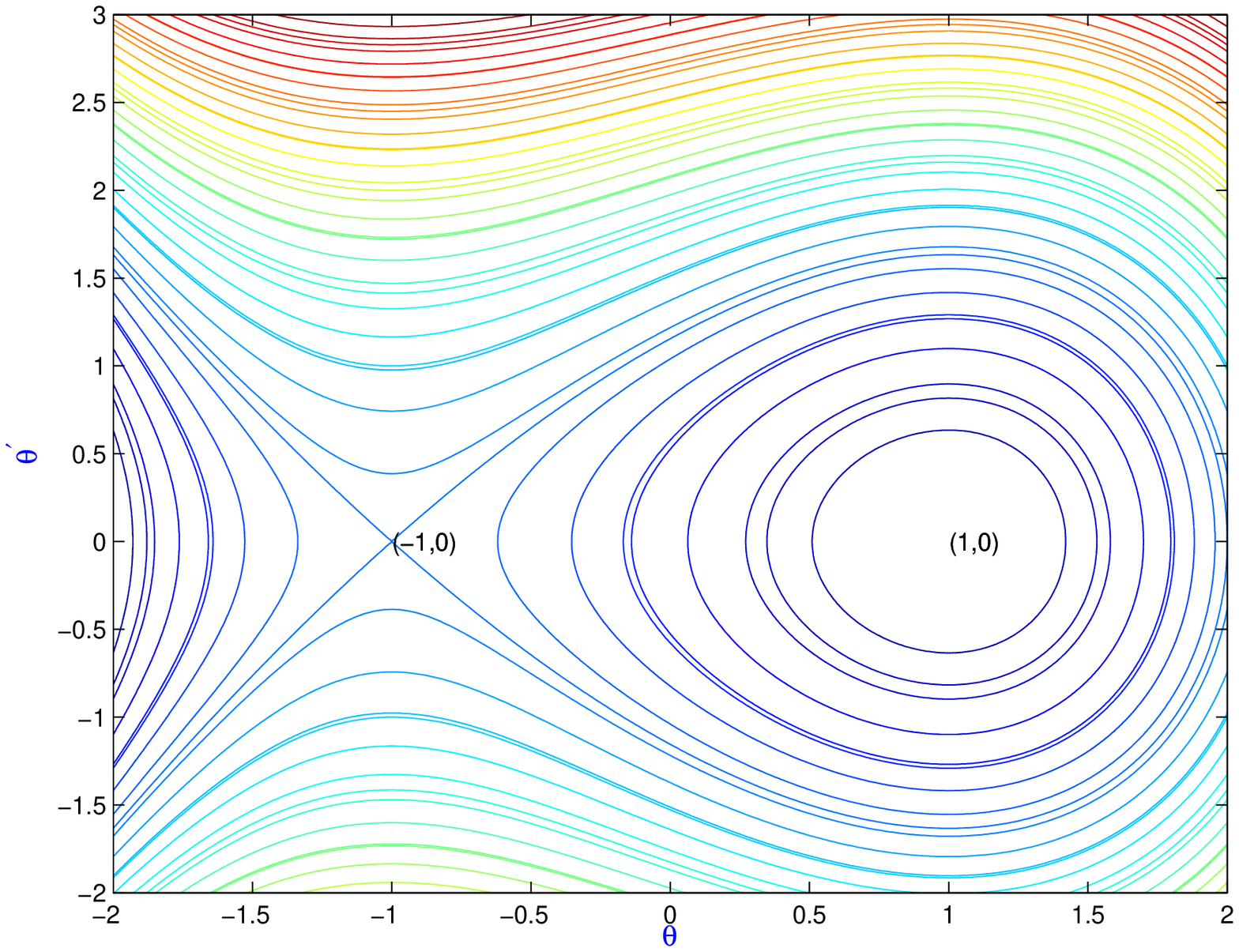}
\end{center}
\begin{center}
{\footnotesize
 {Fig. $3.1$} Classification of  solutions of $\theta^\sec + \theta^2 -1 = 0 $ according to $ \theta(0) $ and $ \theta^\prime(0).$}
\end{center}

 To obtain exact solutions to (3.1),(3.2), we look for ``{\it pseudo-similarity} " solutions  under the
form
\begin{equation}\label{eq:Similarity}
 \psi(x,y) = aF(x,byx^{-1}),
\end{equation}
where $  a = \sqrt{\nu u_\infty} $ and $ b =\sqrt{ \frac{u_\infty}{\nu}}.
$  Assuming $ F(x,t) = f(t) + H(x) $ one sees that (cf. Appendix)

\begin{equation}
\left\{\begin{array}{l}
f^{\sec\prime}+ \t f^\sec+ {f^\prime}^2-1=0,\\
\\
f^\prime(0) = \zeta,\quad f^{\prime}(\infty) = 1,
\end{array}\right.
\end{equation}
and
\[ H(x) = \t\log x + c,\]
where $ \gamma  = v_w\left(\nu u_\infty\right)^{-1/2}$  and $ c $
is constant. Without loss of generality we may take $ c = 0,$ since $\psi(x,y) = \sqrt{\nu u_\infty}f(t) + v_w\log(x)+\sqrt{\nu u_{\infty}} c $ satisfies $(3.1)-(3.2)$ for any real number $c$. Note
that $ \t $
plays the role of suction/injection parameter.\\
To study (3.6) it is  more convenience to consider the second
ordinary differential equation
\begin{equation}
\left\{\begin{array}{l}
\theta^\sec+ \t \theta^\prime+ \theta^2-1=0,\\
\\
\theta(0) = \zeta,\quad \theta(\infty) = 1,
\end{array}\right.
\end{equation}
where $0 \leq \zeta < 1 $ and $ \t\not=0.$ In fact,  the real
number $ \t $ will be taken in $ (0,\infty).$ The existence of
solutions to (3.7) will be proved by means of shooting  method.
Hence, the boundary condition at infinity is replaced by the
condition $ \theta^\prime(0) = d,$ where $ d $ is a real number.
For any $ d $ the new initial--value problem has a unique local
solution $ \theta_d $ defined in the maximal interval of the
existence, say $(0,T_d), T_d \leq \infty.$ We shall see that for
an appropriate $ d $ the solution
 $ \theta_d $ is global and satisfies
\begin{equation} \theta_d(\infty) = 1.\end{equation}
 A simple analysis in the phase plane  shows that  problem (3.6)   may have solutions only for  $\t > 0. $ In fact, the ordinary differential
equation in (3.6) is considered as a nonlinear autonomous system
in $ \mathbb{R}^2, $ with the unknown $(\theta,\theta^\prime),$
mainly
\begin{equation}
\left\{\begin{array}{l}
\theta^\prime=\varphi,\\
\\
\varphi^\prime =-\t \varphi + 1 - \theta^2,\\
\end{array}\right.
\end{equation}
subject to the boundary condition
\begin{equation}
\theta(0) = \zeta,\quad \varphi(0) = d.
\end{equation}
The linear part of the above system at $(1,0)$  is  the matrix
\[ J=\left(\begin{array}{ccc}
                                    0 & 1\\
\\
-2 & -\t
\end{array}\right). \]
The eigenvalues of $ J  $  are
\[ \lambda_1 = \frac{-\t -\sqrt{\t^2-8}}{2},\quad \lambda_2 = \frac{-\t +\sqrt{\t^2-8}}{2},\]
if $ \t \geq \sqrt{8} $ and for $ \vert \t\vert  < \sqrt{8}, $
\[ \lambda_3 = \frac{-\t -i\sqrt{8-\t^2}}{2},\quad \lambda_4 = \frac{-\t +i\sqrt{8-\t^2}}{2}.\]
Therefore, the hyperbolic equilibrium point $(1,0) $ is
asymptotically stable if $ \t $ is positive and unstable for  negative $ \t
$. In particular problem (3.5) has no nontrivial
solutions if $ \t < 0.$ If $ \t > 0 $ we deduce from the above
that there exists $ \delta > 0 $ such that for any $ d $ and $
\zeta $ satisfying $  d^2  + (\zeta - 1)^2 < \delta^2 $ the local
solution $ \theta_d $ is global and satisfies (3.8). In the
following we  construct solution to (3.6) where the condition  $
d^2  + (\zeta - 1)^2 < \delta^2 $ is   not necessarily required.
For a mathematical consideration the parameter $ \zeta $ will be
taken in $(-1,\sqrt{3}].$  The following result  deals with
nonnegative values of $ \zeta.$ Let us consider a real number $ d
$ such that
\begin{equation}
d^2\leq 2\zeta(1-\frac{\zeta^2}{3}),
\end{equation}
where  $ 0 \leq \zeta \leq \sqrt{3}. $ We shall see that any local
solution of (3.9),(3.10) is global and satisfies (3.8). To this
end  we consider again the  Liapunov function $
E(\theta(t),\varphi(t))  = \frac{1}{2}\varphi(t)^2 +
\frac{1}{3}\theta^3-\theta.$ Along an orbit we have
\[ \frac{d}{dt}E(\theta(t),\varphi(t)) = -\t\varphi(t)^2\leq 0.\]
Hence
\[  E(\theta_d(t),\theta_d^\prime(t)) \leq E(\zeta,d),\]
for any  $ t < T_d.$ On the other hand,  from (3.9) and (3.10),
there exists $ t_0 > 0,$ small, such that $ \theta_d $ is positive
on $(0,t_0).$  Assume that $ \theta_d $ vanishes at some $ t_1 >
t_0 $  and suppose that  $ \theta_d^\prime(t_1) \not= 0. $ Because
\[E(\zeta,d) \geq E(\theta_d(t),\theta_d^\prime(t)) \geq \frac{1}{2}\theta_d^\prime(t_1)^2,\]
for all $ 0 \leq t\leq t_1. $ we deduce $ \frac{1}{2}d^2 >
\zeta(1-\frac{1}{3}\zeta^2), $ which contradicts (3.11). Therefore
$ \theta^\prime_d(t_1)= 0.$ In this case we deduce from the
equation of $ \theta_d $ that $ \theta_d^\sec(t_1) = 1 $ and then
$ \theta_d $  is nonnegative on a some neighbourhood of $ t_1.$
Consequently the local solution is nonnegative as long as there
exists.  To show that $ \theta_d $ is global we note that
\[ E(\zeta,d) \geq \frac{1}{2}\theta_d^\prime(t)^2+\frac{1}{3}\theta_d^3(t)-\theta_d(t) \geq -\frac{2}{3},\]
for all $ t \leq T_d$, since $ \theta_d $ is nonnegative. Hence $
\theta_d $ and (then) $ \theta_d^\prime   $ are bounded.
Consequently $ \theta_d $ is global.
It remains to show that $ \theta $ goes to unity at infinity.
To this end we use the {\it Bendixon Criterion}. Let  ${\cal T}$
be the trajectory of $(\theta_d,\theta_d^\prime)$  in  the phase
plane $ (0,\infty)\times\mathbb{R} $ for $ t \geq 0 $ and let
$\Gamma^+({\cal T})$ be its $w$-limit set at infinity.  From the
boudedness of ${\cal T} $  it follows that $ \Gamma^+({\cal T})$
is a nonempty connected and compact subset of
$(0,\infty)\times\mathbb{R}$ ( see, for example \cite[p
226]{Amann}). Moreover $ (-1,0) \not\in \Gamma^+({\cal T}),$ since
$ \theta_d $ is nonnegative.  Note that if $\Gamma^+({\cal T})$
contains the equilibrium point $(1,0) $ then $\Gamma^+({\cal T}) =
\left\{(1,0)\right\},$ since $(1,0)$ is asymptotically stable.
Assume  that $(1,0)\ \not\in\Gamma^+({\cal T}),$
 Applying  the {\it Poincar\'e--Bendixon Theorem }\cite[p 44]{Holmes} we deduce that $\Gamma^+({\cal T})$ is a
cycle, surrounding $(1,0)$. To finish, we shall
prove the nonexistence of such a cycle. We define $
P(\theta,\varphi) = \varphi, Q(\theta,\varphi) = -\t\varphi
+1-\theta^2, \varphi=\theta_d^\prime $ and $ \theta = \theta_d.$
 The function $(\theta,\varphi)$ satisfies the system $ \theta^\prime = P(\theta,\varphi),\ \varphi^\prime = Q(\theta,\varphi).$ Let $ D $ be the bounded
domain of the $(\theta,\varphi)$--plane with boundary ${\Gamma^+}.
$ As $ P $ and $ Q $ are  regular  we deduce, via the {\it
Green--Riemann Theorem},
\begin{equation}
\int\int_D \left(\partial_\varphi Q + \partial_\theta
P\right)d\varphi d\theta=\int_{{\Gamma^+}}\left(Qd\theta - P
d\varphi\right)=0,
\end{equation}
thanks to  the system satisfied by $(\theta,\varphi). $  But $
\partial_\varphi Q + \partial_\theta P = \t $ which is positive.
We get a contradiction.
\\
To complete our analysis, we shall determine a basin of the
critical point $(1,0).$   Let
\[{\cal P} = \left\{(\zeta,d)\ \in \mathbb{R}^2: \zeta > -1, \frac{1}{2}d^2 + \zeta\left(\frac{1}{3}\zeta^2-1\right) < \frac{2}{3} \right\}.\]
Let us consider a one-parameter of family of curves defined by
\[ E(\theta,\varphi) =  \frac{1}{2}\varphi^2 + \frac{1}{3}\theta^3-\theta = C,\]
where $ C $ is a real parameter. Note that, in the phase plane,
this family is  solution curves of system (3.4).  The curve $
\varphi^2 = 2\theta - \frac{2}{3}\theta^3 + \frac{4}{3},$
corresponding to $ C = \frac{2}{3},$ goes through the point
$(2,0)$ and has the saddle $(-1,0) $ ($\t=0$) as its $ \alpha$ and
$w$-limit sets. We note this solution curve by $ {\cal H},$ which
is, in fact, an homoclinic orbit and define a separatrix cycle for
(3.4). We shall see that the bounded open domain with the boundary
$ {\cal H}$ is an attractor set for $(1,0)$ of system (3.9) where
$
 \t > 0.$  This domain is given by $ E(\theta,\varphi) = C, \theta > -1,  $ for all $ -\frac{2}{3} \leq C < \frac{2}{3}, $ which is $ {\cal P}.$ As $
\frac{d}{d t} E \leq 0 $ any solution, with initial data in $
{\cal P} $ cannot leave $  {\cal P}.$  By {\it LaSalle invariance
principle} we deduce that   for any $(\zeta,d) $ in $  {\cal P } $
the $w$-limit set, $ \Gamma^+{(\zeta,d)} $ is a nonempty,
connected subset of $  {\cal P }\cap \left\{\varphi = 0\right\},$
(see \cite[p.  234]{Amann}).   However, if $ \theta\not=1, \varphi = 0
$ is a transversal of the phase--flow, so the $w$--limit set is  $
\left\{(1,0)\right\}.$ This means that  $ {\cal P} $ is a basin
of the critical point $(1,0) $ of (3.9).\\
\begin{center}
\includegraphics[height=2.6 in]{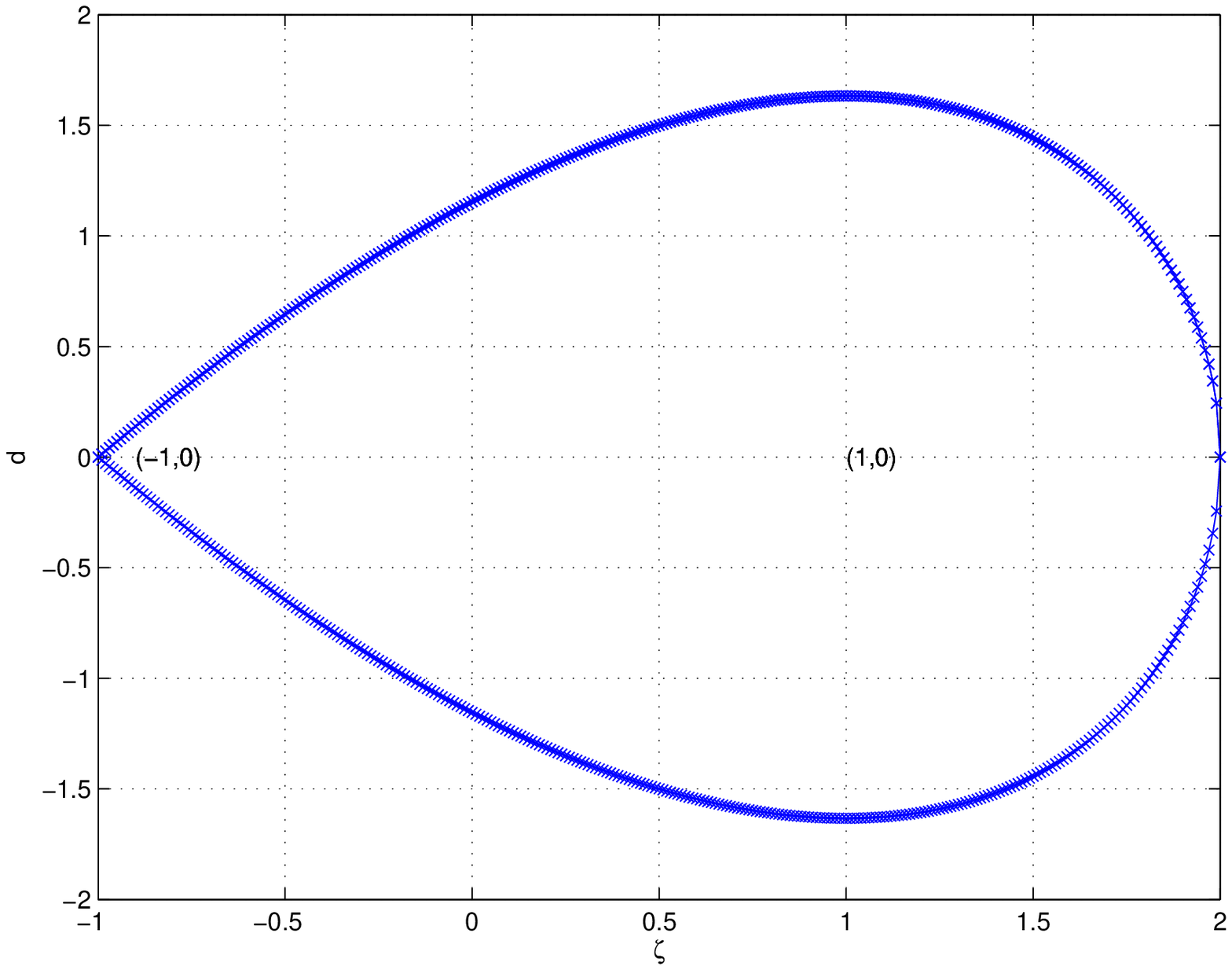}
\begin{center}
{ Fig. $3.2$} A basin of attraction ${\cal P }$ of the critical
point $(1,0)$.
\end{center}
\end{center}

\section{Numerical results}
In this section numerical solutions of the boundary--value problem $(3.7)$ are obtained by using the  fourth-order Runge-Kutta scheme with the shooting method. We plot  the dimensionless velocity $\theta$ in term of the similarity variable $t$, for various value of the shooting parameter $d.$ \\

\bigskip
\begin{center}
\includegraphics[height=2.8in] {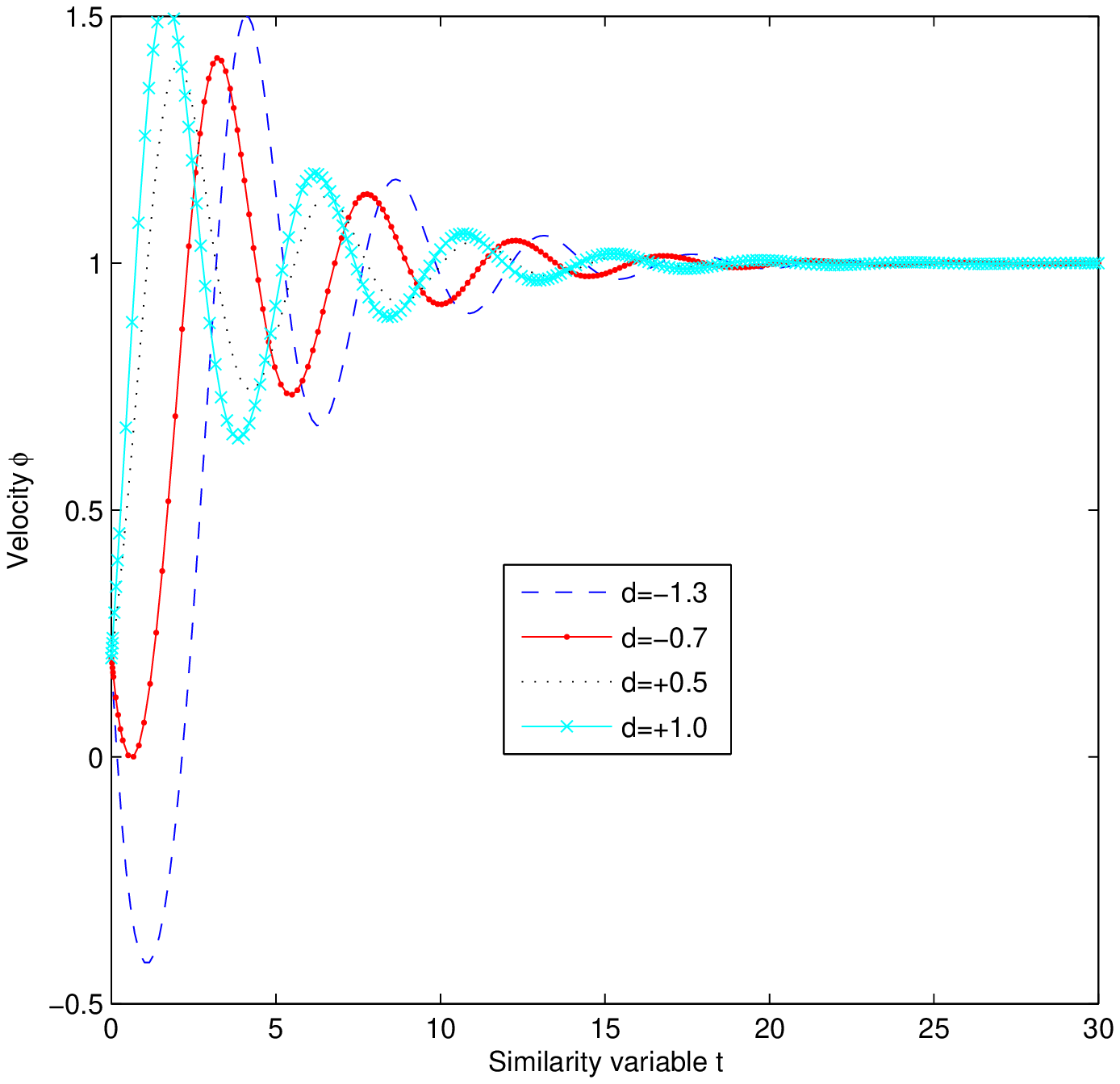}
\begin{center}
{Fig. $4.1$} Velocity profiles  in terms of
$d=\theta^\prime(0)$ for fixed $\zeta =0.2$ and $\gamma =0.5$
\end{center}

\end{center}
\bigskip
\begin{center}
\includegraphics[height=2.8in]{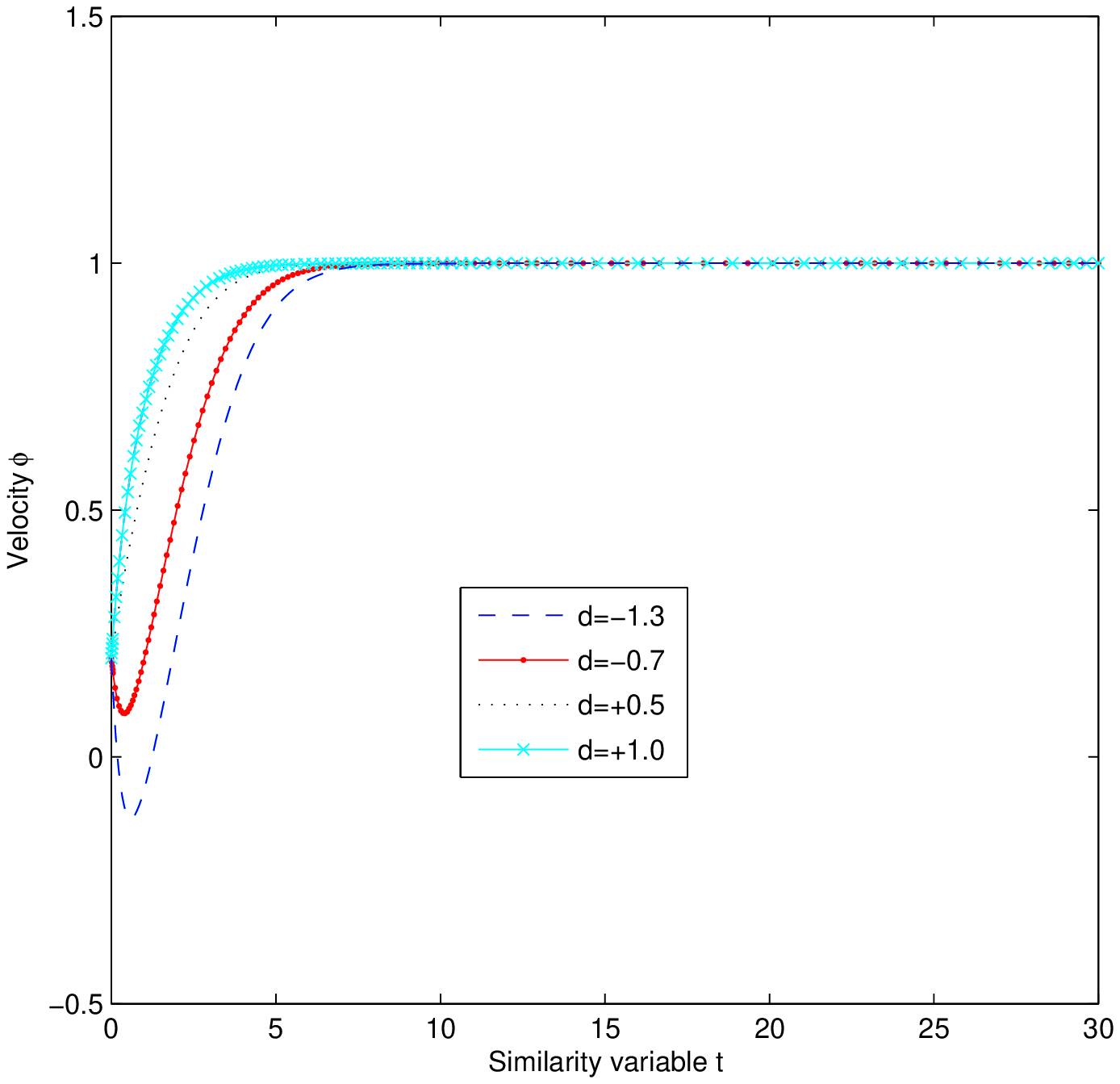}
\begin{center}
{Fig. $4.2$} Velocity profiles in terms of
$d=\theta^\prime(0)$ for fixed $\zeta =0.2$ and $\gamma =2.83$
\end{center}
\end{center}

\section{Conclusion}
\setcounter{equation}{0} \setcounter{theorem}{0}
\setcounter{lemma}{0} \setcounter{remark}{0}
\setcounter{corollary}{0} \qquad In this work the laminar
two-dimensional steady incompressible, boundary layer flow past a
moving  plane  is considered.  It has been shown that the problem
has solutions having a similarity form if the velocity
distribution outside the boundary layer is proportional to $ x^m,$
for some real number $ m.$ In the second part of this paper, we
are interested in question of existence of solutions  in the case
where the external velocity is the inverse-linear function;
$m=-1.$ This situation occurs in the case of sink flow.  To obtain
exact solutions the stream function $ \psi $ is written under the
form
\begin{equation}
\psi(x,y) = \sqrt{\nu u_\infty}f(t) + v_w\log(x).
\end{equation}
  It is shown that
the
  ordinary differential equation satisfied by $ f $  has multiple solutions for any $ v_w $ positive and no solution can exist if $ v_w
\leq 0.$ A sufficient condition for the existence is derived:
\begin{equation}\zeta > -1,\quad \quad \frac{1}{2}f^{\sec}(0)^2 + \zeta\left(\frac{\zeta^2}{3}-1\right) < \frac{2}{3}.\end{equation}  We have
obtained two family of solutions according to  $ \gamma =
v_w\left(\mu u_\infty\right)^{-1/2}. $  If $ \t \geq \sqrt{8} $,
$f^\prime $ is monotonic  and goes to 1 at infinity, but  if $ 0 <
\t < \sqrt{8},$ we have a stable spiral.  The function $f^\prime $
oscillates an infinite number of times and goes to 1. So if we are
interested in solutions to  $(3.7)$ such that
\[ -1 < f^\prime < 1 \] we must take $ u_w, v_w$ and $  u_\infty > 0 $ satisfying $ -u_\infty < u_w < u_\infty $  and $ v_w > (8\nu
u_\infty)^{1/2}.$\\
Condition (4.1) indicates also that for the same positive value of
the suction parameter the permeable wall stretching with velocity
$ u_wx^{-1}, u_w > 0 $ has multiple boundary--layer flows. Every
flow is uniquely determined by the dimensionless skin friction $
f^\sec(0) $ which can be any real  number  in the interval
$\left(-\sqrt{\frac{4}{3} +
2\zeta\left(1-\zeta^2/3\right)},\sqrt{\frac{4}{3} +
2\zeta\left(1-\zeta^2/3\right)}\right),$ where $ \zeta = u_w
u_\infty^{-1}.$ The case $ u_\infty = 0 $ was considered by
Magyari, Pop and Keller \cite{MPK}. The authors showed, by
numerical solutions, that the boundary
layer flow exists only  for a large suction parameter ($\gamma \geq 1.079131$).\\
 The existence of exact solutions of the Falkner-Skan equation under the
present condition was discussed by Rosenhead \cite[pp.
244--246]{Rose} who mentioned that these results may be obtained
by rigorous arguments which, in fact, motivated the present work.
We note, in passing, that it is possible to obtain solutions if
the the skin friction satisfies $ \vert f^\sec(0)\vert >
\sqrt{\frac{4}{3} + 2\zeta\left(1-\zeta^2/3\right)}.$
 \\

\section*{Appendix}
\qquad Let us now  derive problem (3.5). We assume  that  the
external velocity is given by $ u_e(x) = u_\infty x^m,$ where $ m
$ is not necessary equal to -1. We recall that the stream function
satisfies the following equation Ä\begin{equation}
\partial_y\psi\partial_{xy}^2\psi - \partial_{x}\psi\partial_{yy}^2\psi =\nu\partial^3_{yyy}\psi+ mu_\infty^2x^{2m-1},
\end{equation}
with the boundary conditions
\begin{equation}
\partial_y\psi(x,0) = u_w x^{m},\quad \partial_x\psi(x,0) = -v_wx^{\frac{m-1}{2}},\quad \partial_y\psi(x,\infty)=u_\infty x^{m}.
\end{equation}
To obtain exact solutions to (5.3),(5.4), we look for
{\it ``pseudo-similarity " } solutions  under the form
\begin{equation}\label{eq:Similarity}
 \psi(x,y) = ax^{\alpha} F(x,byx^{-\beta}).
\end{equation}
where $\alpha =\frac{m+1}{2},\beta = -\frac{m-1}{2},  a =
\sqrt{\nu u_\infty} $ and $ b =\sqrt{ \frac{u_\infty}{\nu}}. $
   Inserting (4.4) into (4.2),(4.3)  leads to
\begin{equation}
\left\{\begin{array}{l} F^{\sec\prime} +  \frac{1+m}{2} FF^{\sec}-
m({F^\prime}^2-1)+x\left(F^\sec\partial_xF-F^\prime\partial_xF^\prime\right)=0,\\
\\
F^\prime(x,0) = \zeta,\quad F^\prime(x,\infty)=1,\\
\\
\ds \frac{1+m}{2}F(x,0) +x\partial_x F(x,0) = \frac{v_w}{\sqrt{\nu
u_\infty}},
\end{array}\right.
\end{equation}
where the primes denote partial differential with respect to $ t =
\sqrt{\frac{u_\infty}{\nu}}yx^{\frac{m-1}{2}}.$ By writing
\[ F(x,t) = f(t) + H(x),\]
we find
\begin{equation}
\left\{\begin{array}{l}
 f^{\sec\prime} +  \frac{1+m}{2} ff^{\sec}- m\left({f^\prime}^2-1\right)+f^\sec\left(xH^\prime + \frac{1+m}{2}H\right)=0,\\
\\
\ds \frac{1+m}{2}f(0) +\frac{1+m}{2}H(x) + xH^\prime(x)=
\frac{v_w}{\sqrt{\nu u_\infty}},\quad f^\prime(0) = \zeta \in
[0,1),\quad f^{\prime}(\infty) = 1.
\end{array}\right.
\end{equation}
Hence, there exists a real number $ \t $ such that
\begin{equation}
\left\{\begin{array}{l}
f^{\sec\prime} +  \frac{1+m}{2} ff^{\sec}- m\left({f^\prime}^2-1\right)+\t f^\sec=0,\quad t > 0,\\
\\
xH^\prime + \frac{1+m}{2}H = \t, \quad x > 0,\\
\\
\ds \frac{1+m}{2}f(0) +\t= \frac{v_w}{\sqrt{\nu u_\infty}},\quad
f^\prime(0) = \zeta,\quad f^{\prime}(\infty) = 1.
\end{array}\right.
\end{equation}
 First, let us assume that $ m \not=-1.$ The solution $ H $  is given by
\[H(x) = cx^{-\frac{1+m}{2}}+ \frac{2\t}{1+m}, \]
where $ c $ is a constant, and then  $ \psi(x,y) = ac +
ax^{\frac{1+m}{2}}\left(f(t) + \frac{2\t}{1+m}\right).$ The new
function $ g = f + \frac{2\t}{1+m} $ satisfies the Falkner-Skan
equation. Thereafter, we will assume that $ m = -1$ and this leads
to
\begin{equation}
\left\{\begin{array}{l}
f^{\sec\prime}+ \t f^\sec+ {f^\prime}^2-1=0,\\
\\
f^\prime(0) = \zeta,\quad f^{\prime}(\infty) = 1,\ \t=
\frac{v_w}{\sqrt{\nu u_\infty}},
\end{array}\right.
\end{equation}
and
\[ H(x) = \t\log x + c,\quad c = const.\]
\begin{quote} {\bf Acknowledgments. }{\small The authors would like to
thank  R. Kersner for   stimulating discussions and  the referees
for their careful  reading of the original manuscript and for
making constructive suggestions, which have improved the
presentation of this work. This work was partially supported   by
Direction des Affaires Internationales  (UPJV) Amiens, France  and  by PAI No MA/05/116. }
\end{quote}

\end{document}